\documentclass[journal]{IEEEtran}

\usepackage{amssymb}
\usepackage{amsthm}
\usepackage{graphicx}
\usepackage{array,color}
\usepackage{amsmath}
\usepackage{stfloats}
\usepackage{graphicx}
\usepackage{subfigure}
\usepackage{tabularx}
\usepackage{epsfig,epsf,color,balance,cite}
\usepackage{algorithmic}
\usepackage{algorithm}
\usepackage{bm}
\usepackage{textcomp}
\usepackage{color}
\usepackage{multirow}
\usepackage{cite}
\usepackage{enumerate}
\usepackage{cases}
\usepackage{color}
\usepackage{url}
\usepackage{epstopdf}

\begin{document}
	\title{Sum Rate Maximization for Intelligent Reflecting Surface Assisted Terahertz Communications}
	\author{\IEEEauthorblockN{Yijin Pan, Kezhi Wang, Cunhua Pan, Huiling Zhu and Jiangzhou Wang,~\IEEEmembership{Fellow,~IEEE} \vspace{-1em}}
	\thanks{Y. Pan are with the National Mobile Communications Research Laboratory, Southeast University, Nanjing 211111, China. She is also with School of Engineering and Digital Arts, University of Kent, UK. Email: panyj@seu.edu.cn, y.pan@kent.ac.uk.}
	\thanks{K. Wang is with the Department of Computer and Information Sciences, Northumbria University, UK. Email: kezhi.wang@northumbria.ac.uk.}
	\thanks{C. Pan is with the School of Electronic Engineering and Computer Science, Queen Mary, University of London, UK. Email: c.pan@qmul.ac.uk.}
	\thanks{J. Wang and H. Zhu are with the School of Engineering and Digital Arts, University of Kent, UK. Email: J.Z.Wang@kent.ac.uk, H.Zhu@kent.ac.uk.}}
	
	\maketitle

	\begin{abstract}
	In this paper, an intelligent reflecting surface (IRS) is deployed to assist the terahertz (THz) communications. The molecular absorption causes path loss peaks to appear in the THz frequency band, and the fading peak is greatly affected by the transmission distance. In this paper, we aim to maximize the sum rate with individual rate constraints, in which the IRS location, IRS phase shift, the allocation of sub-bands of the THz spectrum, and power control for UEs are jointly optimized. For the special case of a single user equipment (UE) with a single sub-band, the globally optimal solution is provided. For the general case with multiple UEs, the block coordinate searching (BCS) based algorithm is proposed to solve the non-convex problem. Simulation results show that the proposed scheme can significantly enhance system performance.
	\end{abstract}
	

	\section{Introduction}
	
	The terahertz (THz) band wireless transmission has been envisioned as a promising solution to meet the ultra-high data rate requirements of the emerging applications such as the virtual reality (VR) service.
	However, due to its ultra-high frequency, the propagation at THz covers short-range area and is susceptible to blockages.
	This issue becomes more pronounced in the indoor scenario with furniture and complex interior structure\cite{Chaccour.2020b}.
	Recently, intelligent reflecting surface (IRS), also known as reconfigurable intelligent surface (RIS), has been widely proposed to reconfigure wireless propagation environment  to enhance the system performance, such as the simultaneous wireless information and power transfer (SWIPT) system \cite{9110849}, {orthogonal frequency division multiple access (OFDMA) system\cite{Yang.2019} }and multicell network \cite{9090356} through careful design of the phase shifts of the IRS.
	Due to its capability of constructing an alternative non-line-of-sight (NLoS) communication link, the IRS is very attractive for the applications in THz communications that are sensitive to blockages. 
	However, the study on IRS-aided THz communications is still in its infancy  \cite{Chen.20198112019813,Ma.2020,Ning.2019} and numerous practical design issues are not yet addressed.

	The sum rate performance of the IRS-aided THz communication system was maximized by optimizing the phase shift of the IRS\cite{Chen.20198112019813}.
	The IRS-assisted massive multiple input multiple output (MIMO) transmissions in THz band were investigated in \cite{Ma.2020,Ning.2019}. 
	Although the above-mentioned literature studied the IRS-aided THz communications, the special features associated with the THz band were not considered.
	In fact, the bandwidth offered by the THz Band ranges from 0.1 THz to several THz.
	Due to the propagation loss and high molecular absorption in THz band, severe path loss peaks appear in different frequencies.
	Hence, the total bandwidth is divided into several sub-bands with different bandwidths\cite{Jornet.2011}.
	In addition, the path loss of the sub-band varies with the communication distance\cite{Han.2016,Han.2015}.
	Thus, the frequency and distance-dependent sub-bands in the THz communications should be taken into consideration in the system design.

	However, it is challenging to utilize multiple sub-bands in THz communications, as it needs to be intelligently selected to avoid the path loss peaks.
	Meanwhile, the peaks of path loss in the THz band also depend on the transmission distance, which highly relies on the location of IRS in the NLoS links.
	Fortunately, in indoor applications, the locations of UEs can be predicted by using the historical statistical information of users' movement and the layout of furniture, and this information can be exploited to decide the optimal deployment of the IRS during wireless network design.
	Therefore, it is imperative to jointly optimize the deployment of IRS, the reflecting phase shift, along with the sub-band allocation to enhance the IRS-assisted THz transmission.
	Unfortunately, this issue has not yet been addressed in the existing literature.

	Against the above background, in this paper, the THz transmission in a blocked indoor scenario is studied with the assistance of an IRS.  
	The THz wireless channel consists of a set of frequency and distance-dependent sub-bands, and the channel fading in each sub-band is affected by both spreading loss and absorbent molecules.
	Our target is to maximize the summation of the achievable rates of UEs by optimizing the IRS location, the IRS phase shift, sub-band allocation, and power control.
	The globally optimal solution is provided for the special case of single UE with a single sub-band, and a block coordinate searching (BCS) algorithm is proposed for the general case with multiple UEs.
	Simulation results are also provided for performance evaluation.

	\section{System Model}
\begin{figure}
		\centering
		\vspace{-1em}
		\includegraphics[width=0.35\textwidth]{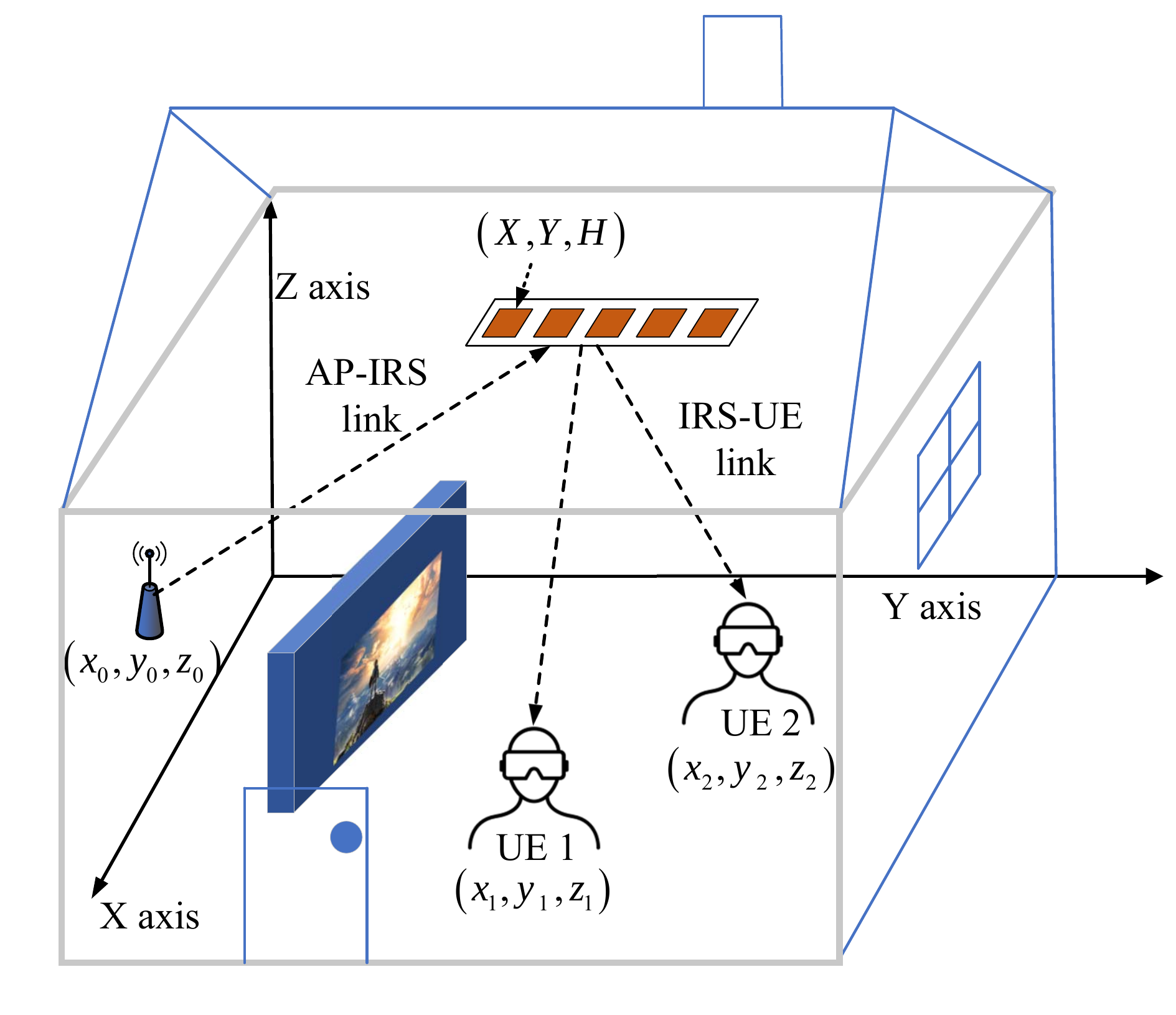}
		\vspace{-1em}
		\caption{The IRS-assisted THz transmission scenario.}
		\vspace{-1em}
		\label{fig1}
	\end{figure}

	\newtheorem{proposition}{\textbf{Proposition}}
	Consider the downlink transmission of an access point (AP) operating in THz frequency to support the VR service in an indoor scenario as shown in Fig. \ref{fig1}.
	The LoS link from the AP to a given area may be blocked by obstacles such as the pillar or wall, and this area is assumed to be rectangular with length $L$ and width $W$. 
	Then, the set $\mathcal{U}$ of $U$ UEs in this area are served by the AP with the aid of an IRS as shown in Fig. \ref{fig1}, in which the IRS has $N$ passive reflecting units.
	Suppose that the reflecting coefficients of all units share the same amplitude value of one and have different phase shifts.
	Let $\phi_{n}$ denote the phase shift of the $n$-th reflecting unit of the IRS, which can be carefully adjusted by an IRS controller.
	Then, the reflecting coefficient of the $n$-th reflecting unit is $\Gamma_{n} =  e^{j\phi_{n}}$.
	
	Assume that the IRS can be installed on the ceiling with height $H$ and parallel to the Y-axis, and the location of the first reflecting unit is denoted by $\bm{l}_1 = [X, Y, H]^T$.
	The separation between adjacent IRS elements is denoted by $\Delta$, so that the location of the $n$-th reflecting unit is $\bm{l}_n = [X,Y+(n-1)\Delta,H]^T$.
	Furthermore, the location of AP is denoted by $\bm{w}_0 = [x_0, y_0,z_0]^T$.
	The location of UE $u$ is denoted by $\bm{w}_u = [x_u, y_u,z_u]^T$, {which can be estimated by using the information of furniture layout and the statistics of users' historical activities.}
	
	Assume that the IRS is a uniform linear array, so that the following steering vectors are introduced to help represent the channel gain. 
	For the link from the AP to the IRS, we define steering vector $\bm{e}^t(f,\bm{l})=\left [1, e^{-j \theta_1(f,\bm{l})}, \cdots,e^{-j \theta_N(f,\bm{l})}\right ]$, where $\bm{l} = [X,Y]$.
	The phase $\theta_n(f,\bm{l})$ represents the phase difference of the incoming signal to the $n$-th reflecting unit relative to the first unit with transmission frequency $f$, and it can then be calculated as
	\begin{equation}
	\theta_n(f,\bm{l}) = \frac{2\pi f }{c} \frac{\bm{r}_0^T}{|\bm{r}_0|}(\bm{l}_n-\bm{l}_1) 
	= \frac{2\pi f }{c} \frac{(Y-y_0)(n-1)\Delta}{|\bm{r}_0|},
	\end{equation}
	where $|\bm{x}|$ represents the Euclidean norm of vector $\bm x$,  and ${\bm{r}_0} = [X-x_0, Y-y_0,H-z_0]^T$.
	
	For the link from the IRS to UE $u$, we define the steering vector $\bm{e}_{u}^r(f,\bm{l})= \left[1, e^{-j \vartheta_1(f,\bm{l})}, \cdots,e^{-j \vartheta_N(f,\bm{l})}\right]$. 
	Then, the phase difference of the signal received at UE $u$ from the $n$-th reflecting unit relative to the first unit with transmission frequency $f$ is 
	\begin{equation}
	\vartheta^u_n(f,\bm{l}) = \frac{2\pi f }{c} \frac{\bm{r}_u^T}{|\bm{r}_u|}(\bm{l}_n-\bm{l}_1) 
	= \frac{2\pi f }{c} \frac{(y_u-Y)(n-1)\Delta}{|\bm{r}_u|},
	\end{equation}
	where ${\bm{r}_u} = [x_u-X, y_u-Y,z_u-H]^T$, $u=1,\cdots,U$.

\begin{figure}
		\centering
		\vspace{-1em}
		\includegraphics[width=0.35\textwidth]{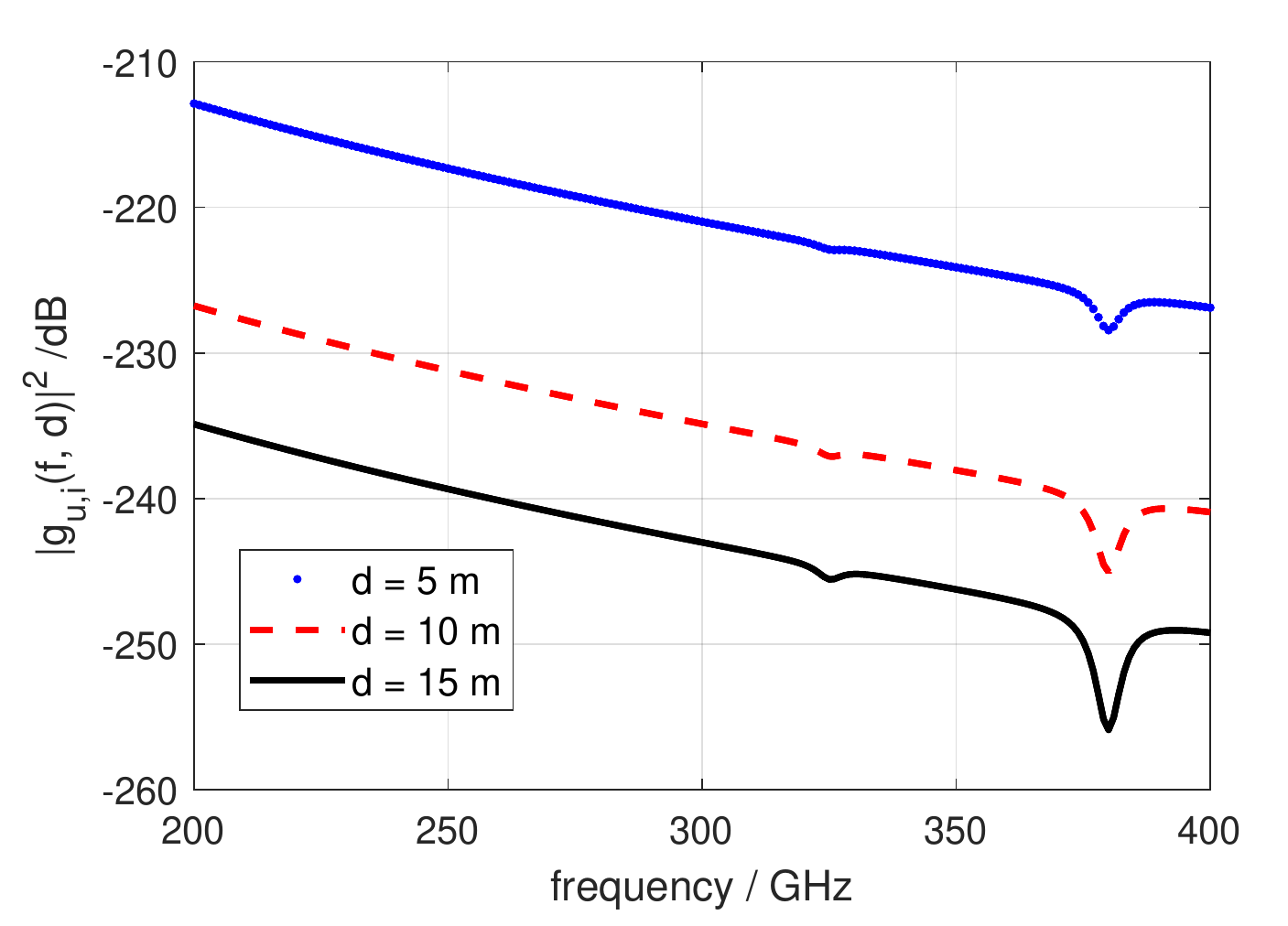}
		\vspace{-1em}
		\caption{The reflected channel gain $|g_{u,i}(f_i, d_{u})|^2$ with the atmospheric pressure $101325$ Pa, $50\%$ relative humidity, and 23$^{\circ}$C .}
		\vspace{-1em}
		\label{fig3}
	\end{figure}

	
	In THz band, apart from the spreading loss affected by the frequency and distance, the absorbent molecules composited in the transmission medium cause several peaks of channel attenuation, as shown in Fig. \ref{fig3}.
	As a consequence, the total bandwidth of THz band needs to be divided into several sub-bands, and the channel capacity is calculated by adding up the rates of all sub-bands. 
	Let $f_i$ denote the central frequency of the $i$-th sub-band, the set of total sub-bands is denoted by $\mathcal{I}$, and the total number of sub-bands is assumed to be $I$.
	According to \cite{Han.2015,Tang.2019}, {in the near field scenario,} the cascaded NLoS channel gain of the AP-IRS-UE $u$ link on the $i$-th sub-band is 
	\begin{equation}
	g_{u,i}(f_i, d_{u})= \left( \frac{c}{4 \pi f_i d_{u} } \right) e^{-j 2\pi f_i \frac{d_{u}}{c}} e^{-\frac{1}{2}K(f_i)d_{u}},
	\end{equation}
	where $d_u = |\bm{r}_0|+ |\bm{r}_u|$, $K(f_i)$ is the overall absorption coefficient of the medium, and $c$ is the  light speed.
	Then, the reflecting channel of AP-IRS-UE $u$ link can be expressed as 
	\begin{equation}
	h_{u,i}(f_i,\bm{\Phi}, \bm{l})
	=  g_{u,i}(f_i, d_{u})\bm{e}^t(f_i,\bm{l})\bm{\Phi}{\bm{e}^r_u(f_i,\bm{l})}^T,
	\end{equation}
	where $\bm \Phi= \text{diag} (\Gamma_{1}, \cdots,\Gamma_{N})$.
	
	To evaluate $K(f_i)$, we adopt a simplified molecular absorption coefficient model for 200 – 400 GHz frequency band\cite{Boulogeorgos.20186252018628}, which has two major absorption peaks at about 325 GHz and 380 GHz. 
	This simplified model only depends on the volume mixing ratio of water (humidity) $\mu_{w}$ and frequency $f$ (Hz):
	\begin{multline}
	\!\!\!\!\!\!K(f) \!= \!\frac{A(\mu_{w})}{B(\mu_{w}) + \left(\frac{f}{100c}\!\!-\!\!10.835\right)}
	+ \frac{C(\mu_{w})}{D(\mu_{w}) + \left(\frac{f}{100c}\!-\!12.664\right)} \\
	+p_1 f^3 + p_2 f^2 + p_3 f+ p_4,
	\end{multline}
	where $A(\mu_{w})=0.2205\mu_{w}(0.1303\mu_{w}+0.0294)$, $B(\mu_{w})=(0.4093\mu_{w}+0.0925)^2$, $C(\mu_{w})=2.014\mu_{w}(0.1702\mu_{w}+0.0303)$, 
	$D(\mu_{w})= (0.537 \mu_{w}+0.0956)^2$, 
	$p_1 =5.54 \times 10^{-37}\text{Hz}^{-3}$, 
	$p_2 = -3.94 \times  10^{-25} \text{Hz}^{-2}$,
	$p_3 = 9.06 \times 10^{-14} \text{Hz}^{-1}$, and
	$p_4 = -6.36 \times 10^{-3}$.
	The volume mixing ratio of water vapour $\mu_{w}$ is evaluated as
	\begin{equation}
	\mu_{w} = \frac{\phi_{H}}{100} \frac{p_w(T,p_{ss})}{p_{ss}},
	\end{equation}
	where $\phi_{H}$ and $p_{ss}$ (measured in hectopascal) respectively represent the relative humidity and the pressure. 
	The saturated water vapour partial pressure $p_w(T,p_{ss})$ also depends on temperature $T$ (measured in $^{\circ}$C), according to Buck equation \cite{Alduchov.1996}, which is calculated as 
	$p_w(T, p_{ss})= 6.1121(1.0007 +3.46\times 10^{-8}p_{ss})\exp\left(\frac{17.502T}{240.97+T}\right)$.
	
	Similar to \cite{Han.2015}, we assume that each sub-band is less than coherence bandwidth, and the separation between adjacent sub-bands is assumed to be sufficiently large.
	Therefore, the inter-symbol interference (ISI) and interband interference (IBI) can be eliminated and narrowband communication on each  sub-band can be achieved\cite{Du.2020}. 
	Let $B_i$ denote the bandwidth of the $i$-th sub-band, and the power spectral density of the noise in the $i$-th sub-band is expressed as $S_N(f_i)$.
	The obtained transmission rate of UE $u$ on the $i$-th sub-band is given by 
	\begin{equation}
	R_{u,i}(f_i, p^t_i,\bm{\Phi},\bm{l}) =B_i  \log \left( 1 + \frac{p^t_{i}|h_{u,i}(f_i,\bm{\Phi}, \bm{l})|^2 }{S_N(f_i) B_i}\right),
	\end{equation}
	where $p^t_{i}$  is the transmit power on the $i$-th sub-band.

	\section{Problem Formulation}
	
	Let the binary variable $\alpha_{u,i}$ indicate that the signal is transmitted to the UE $u$ on the $i$-th sub-band. Assume that each sub-band is only allocated to one UE. Then, we have the following constraint as
	\begin{equation}
	C1: \alpha_{u,i} \in \{0,1\}, \sum_{u =1}^U \alpha_{u,i} = 1, i \in \mathcal{I}, u \in \mathcal{U}. 
	\end{equation}
	The transmit power $p^t_{i}$ is constrained by AP's maximum power $p_{max} $ as
	\begin{equation}
	C2: \sum_{i =1}^I p^t_{i} \leq p_{max}.
	\end{equation}
	In addition, as the phase shift of each reflecting element of the IRS varies within the range of $[0, 2\pi]$, the following constraint is introduced on the IRS phase coefficient matrix $\bm{\Phi}$ as
	\begin{equation} \label{C1st}
	C3: |\Gamma_{n}|^2 =1, 1\leq n \leq N. 
	\end{equation}
	Furthermore, we have the following restriction on the deployment of IRS location $\bm l = [X,Y]$ as
	\begin{equation}
	C4:   0 < X < W, 0 < Y < L. 
	\end{equation}
	Then, to serve multiple UEs, the rate requirement should be satisfied for each UE as
	\begin{equation}
	C5:  \sum_{i =1}^I \alpha_{u,i}R_{u,i}(f_i, p^t_i,\bm{\Phi},\bm{l}) \geq R_u^{th}, \forall u \in \mathcal{U}.
	\end{equation}
	
	Based on the above descriptions, we aim to maximize the sum rate of the IRS-aided THz system by jointly optimizing the IRS location $\bm{l}$, sub-bands allocation $\{\alpha_{u,i}\}$, the transmit power $\{p^t_i\}$ and the phase shift of IRS $\{\bm{\Phi}\}$.
	Then, we formulate the following optimization problem as
	\begin{subequations}\label{pro1}
		\begin{align}
		\underset{\{\bm{\Phi}\},\{\alpha_{u,i}\}, \{p^t_i\}, \bm{l}}{\text{max}}  
		&\quad R =  \sum_{u \in \mathcal{U}}\sum_{i  \in \mathcal{I}} \alpha_{u,i}R_{u,i}(f_i, p^t_i,\bm{\Phi},\bm{l}) \label{obj1}  \\
		\text{s.t.} & \quad C1-C5.
		\end{align}
	\end{subequations}
	However, Problem (\ref{pro1}) is challenging to solve due to the complex form of $|h_{u,i}(f_i,\bm{\Phi}, \bm{l})|^2$.
	Therefore, in the following, we first present the optimal solution for the special case of single UE, and then we propose an iterative BCS algorithm to solve Problem (\ref{pro1}) for the multiple-UE case.
	
	%
	%
	%
	%
	
	\section{Single UE Single Sub-band Case}
	
	Consider that only one UE is located in the service area.
	It is assumed that the largest flat sub-band with central frequency $f_0$ is utilized, and all the power is allocated to this UE on the sub-band, i.e., $p_0^t = p_{max}$.
	The resulting channel gain is expressed as $|h_{u,i}(f_0,\bm{\Phi}, \bm{l})|^2 $, then the IRS phase shift is readily obtained as
	\begin{equation}
	\phi_{n}^* = \frac{2\pi f_0 (n-1)\Delta }{c} \left( \frac{(y_u-Y)}{|\bm{r}_u|} + \frac{(Y-y_0)}{|\bm{r}_0|} \right) .
	\end{equation}
	Then, the channel gain  is expressed as $|h_{u,i}(f_0,\bm{\Phi}^*, \bm{l})|^2 = \left( \frac{c}{4 \pi f_0 d_{u} } \right)^2 e^{-K(f_0)d_{u}}N^2$.
	It is observed that $|h_{u,i}(f_0,\bm{\Phi}^*, \bm{l})|^2 $ decreases with $d_{u}$, so that the optimal IRS location $(X,Y,H)$ is determined by solving the following problem:
	\begin{equation}\label{pro1_1}
		\underset{0 < X < W, 0 < Y < L}{\text{min}}d_u = D_0(X,Y) +D_u(X,Y)
	\end{equation}
	where $D_k(X,Y) = \sqrt{(X\!-\!x_k)^2+(Y\!-\!y_k)^2+ H_k} , k = 0,u$, $H_u = (H-z_u)^2$, and $H_0 = (H-z_0)^2$.
	Then, the Hessian matrix of $D_k(X,Y)$ denoted by $\bm{H}_{D_k}$ is given by
	\begin{equation}
	\bm{H}_{D_k}\!=\!D_k(X,Y)^{-\frac{3}{2}}
	\left[\begin{matrix}
	(Y\!-\!y_k)^2+ H_k\!\!\! &\!\! \!\!-(X\!-\!x_k)(Y\!-\!y_k)   \\
	-(X\!-\!x_k)(Y\!-\!y_k)\! \!\!& \!\!\!\!  (X\!-\!x_k)^2+ H_k
	\end{matrix}\right]  . \nonumber
	\end{equation}
	It is verified that the determinant of $\bm{H}_{D_k}$ is non-negative, so that $\bm{H}_{D_k}$ is semi-positive definite.
	As a result, Problem (\ref{pro1_1}) is a convex problem, and it can be solved by the standard algorithms, such as the interior point method.
	
	\section{Multiple UEs Multiple Sub-bands Case}
	In this section, we consider the more general case with multiple sub-bands and multiple UEs.
	
	Defining $\delta_i = \frac{1}{S_N(f_i) B_i}$ and introducing the auxiliary variables $\{t_{u,i}\}$, we reformulate Problem (\ref{pro1}) as:
	\begin{subequations}\label{pro2}
		\begin{align}
		\!\!\!\underset{ \underset{\{\alpha_{u,i}\}, \{p^t_i\}, }{\{\bm{\Phi}\},\bm{l}, \{t_{u,i}\}}}{\text{max}}
		& R_s(t_{u,i})=\sum_{u \in \mathcal{U}}\sum_{i  \in \mathcal{I}} \alpha_{u,i} B_i  \log \left( 1 + \delta_i t_{u,i}\right) \label{obj2}  \\
		\text{s.t.} & \ p^t_i|h_{u,i}(f_i,\bm{\Phi}, \bm{l})|^2\!\! \geq \! \alpha_{u,i} t_{u,i},\forall u \in \mathcal{U}, i \in  \mathcal{I}, \label{P2_st1} \\
		& \ \sum_{i \in \mathcal{I}}B_i \alpha_{u,i} \log \left( 1 + \delta_i t_{u,i} \right) \geq R_u^{th}, \forall u \in \mathcal{U}, \label{P2_st2}  \\
		& \ t_{u,i}\geq 0,\forall u \in \mathcal{U}, i \in  \mathcal{I},  C1-C4. \nonumber
		\end{align}
	\end{subequations}

	To solve Problem (\ref{pro2}),  we first fix the location $\bm{l}$ of the IRS, and optimize the IRS phase shift and the sub-bands/power allocation. 
	As the expression of $|h_{u,i}(f_i,\bm{\Phi}, \bm{l})|^2$ is intractable, the two-dimensional search is then adopted to find the optimal location of IRS.
	
	\subsection{IRS phase Shift Optimization}
	
	With given $\bm{l}$, $\{t_{u,i}\}$, $\{\alpha_{u,i}\}$ and $\{p_i^t\}$, (\ref{P2_st1}) can be represented as
	\begin{equation}
	p^t_i|g_{u,i}(f_i, d_{u}) \bm{e}^t(f_i,\bm{l})\bm{\Phi}{\bm{e}^r_u(f_i,\bm{l})}^T|^2 \geq  t_{u,i},\forall u \in \mathcal{U}, i \in \mathcal{I}_u,
	\end{equation}
	where $\mathcal{I}_u$ represents the set of sub-bands that are allocated to UE $u$.
	Then, we define
	\begin{multline} \nonumber
	\bm{e}_{u} (f_i) =\sqrt{p^t_i} g_{u,i}(f_i, d_{u})[1,  e^{-j(\theta_1(f_i) + \vartheta_1(f_i))}, \cdots, \\ e^{-j (\theta_{N-1}(f_i) + \vartheta_{N-1}(f_i))}], \bm{\phi} = [\Gamma_{1}, \cdots, \Gamma_{N}]^T.
	\end{multline}
	To optimize the IRS phase shift vector $\bm{\phi}$, Problem (\ref{pro2})  is simplified to
	\begin{subequations}\label{pro4}
		\begin{align}
		\underset{\bm{\phi}}{\text{max}}  
		&\quad R_s(t_{u,i}) \label{obj4}  \\
		\text{s.t.} & \quad |\Gamma_{n}|^2 =1, 1\leq n \leq N, \label{P4_st1} \\
		& \quad |\bm{e}_{u} (f_i)\bm{\phi}|^2 \geq t_{u,i} , u \in \mathcal{U}, i \in \mathcal{I}_u. \label{P4_st2}
		\end{align}
	\end{subequations}
	As the feasible set of Problem (\ref{pro4}) is non-convex, Problem (\ref{pro4}) is difficult to solve in this original formulation.
	Define $w_{u,i} = \bm{e}_{u} (f_i)\bm{\phi}$. Then, $T_{u,i}(w_{u,i}) = w_{u,i} ^2$  is convex with respect to $w_{u,i}$.
	Thus, its lower bound surrogate function at $\hat{w}_{u,i}$ could be obtained by the first-order Taylor approximation as
	\begin{multline}\label{Tdiv}
	T_{u,i}(w_{u,i})  \geq T_{u,i}(\hat{w}_{u,i}) 
	+ \triangledown_{w_{u,i}} T_{u,i}|_{w_{u,i} = \hat{w}_{u,i}}(w_{u,i}-\hat{w}_{u,i})\\
	+ \triangledown_{w_{u,i}^*} T_{u,i}|_{w_{u,i}=\hat{w}_{u,i}}(w_{u,i}^*-\hat{w}_{u,i}^*).
	\end{multline}
	Substituting $w_{u,i} = \bm{e}_{u} (f_i)\bm{\phi}$ and $\hat{w}_{u,i} = \bm{e}_{u} (f_i)\bm{\hat{\phi}}$ into the right hand side of (\ref{Tdiv}), we have
	\begin{equation}\label{TdivFun}
	\!\!\! T_{u,i}(\bm{e}_{u} (f_i)\bm{\phi}) \geq  2 \Re\{\bm{e}_{u} (f_i)\bm{\hat{\phi}}^H \bm{e}_{u} (f_i)^H\bm{\phi}\} - |\bm{e}_{u} (f_i)\bm{\hat{\phi}}|^2. 
	\end{equation}
	Then, Problem (\ref{pro4}) can be addressed by solving a sequence of simpler problems.
	Based on (\ref{TdivFun}), for given $\bm{\hat{\phi}}$, the convex approximation problem of Problem (\ref{pro4}) can be constructed as
	\begin{subequations}\label{pro5}
		\begin{align}
		\underset{\bm{\phi}}{\text{max}}  
		&\ R_s(t_{u,i}) \label{obj5}  \\
		\text{s.t.} & \ |\Gamma_{n}|^2 =1, 1\leq n \leq N, \label{P5_st1} \\
		& \ 2 \Re\{\bm \Theta_{u,i}(\bm{\hat{\phi}})\bm{\phi}\}  \geq   \Psi_{u,i}(\bm{\hat{\phi}}) + t_{u,i} , u \in \mathcal{U}, i \in \mathcal{I}_u, \label{P5_st2}
		\end{align}
	\end{subequations}
	where 
	$\Psi_{u,i}(\bm{\hat{\phi}}) =|\bm{e}_{u} (f_i)\bm{\hat{\phi}}|^2$, and
	$\bm \Theta_{u,i}(\bm{\hat{\phi}}) = \bm{e}_{u} (f_i)\bm{\hat{\phi}}^H \bm{e}_{u} (f_i)^H $.
	However, as the constraint (\ref{P5_st1}) is non-convex, the Lagrangian dual method cannot be applied to solve Problem (\ref{pro5}) due to the non-zero dual gap.
	In the following, we adopt a pricing mechanism to solve Problem (\ref{pro5}), where a series of non-negative prices $\{\rho_{u,i}\}$ are introduced in  constraints (\ref{P5_st2}).
	Then, a penalty term is introduced to the objective function, and the problem is transformed to
	\begin{multline}\label{pro6}
		\underset{|\Gamma_{n}|^2 =1}{\text{max}}  
		\ R_s(t_{u,i}) +\sum_{u \in \mathcal{U}}\sum_{i  \in \mathcal{I}_u}\rho_{u,i}\left(2 \Re\{\bm \Theta_{u,i}(\bm{\hat{\phi}})\bm{\phi}\}  \right.\\ - \left. \Psi_{u,i}(\bm{\hat{\phi}})-t_{u,i}\right).
	\end{multline}
	With given $\{\rho_{u,i}\}$, the optimal solution $\bm{\phi}$ to Problem (\ref{pro6}) is given by 
	\begin{equation}\label{Phi}
	{\phi}^*_{n}= {\hat{\bm{\Theta}}_n(\rho_{u,i})}, \text{and } \hat{\bm{\Theta}}(\rho_{u,i}) = \arg \left(\sum_{u \in \mathcal{U}}\sum_{i  \in \mathcal{I}_u} 2\rho_{u,i} \bm \Theta_{u,i}(\bm{\hat{\phi}})\right) ,
	\end{equation}
	where $\hat{\bm{\Theta}}_n(\rho_{u,i})$ is the $n$-th element of $\hat{\bm{\Theta}}(\rho_{u,i})$.
	
	If the obtained solution $\bm{\phi}^*$ is not feasible, the introduced penalty term will decrease the objective value.
	Consequently, the pricing factors $\{\rho_{u,i}\}$ should be optimized so that the introduced penalty term  $\rho_{u,i}\left(2 \Re\{\bm \Theta_{u,i}(\bm{\hat{\phi}})\bm{\phi}\}  - \Psi_{u,i}(\bm{\hat{\phi}}) - t_{u,i}\right)$ is minimized.
	Then, to obtain pricing factor $\rho_{u,i}^*$, we employ the sub-gradient descent based method.
	To be specific, $\{\rho_{u,i}^{(t)}\}$ in the $t$-th iteration is updated as
	\begin{align}
	\!\!\!\!\!\rho_{u,i}^{(t)} \!=\! \left[\rho_{u,i}^{(t-1)}\!\! - \!\!\tau_{u,i}^{(t)}\left(2 \Re\{\bm \Theta_{u,i}(\bm{\hat{\phi}})\bm{\phi}\} \! -\! \Psi_{u,i}(\bm{\hat{\phi}}) \!-\! t_{u,i}\!\right)\!\right]^+ \!\!\!,\label{rhokt} 
	\end{align}
	where $[a]^+ = \max\{0,a\}$, $\tau_{u,i}^{(t)}$ is the positive step-size in the $t$-th iteration.

	\begin{algorithm}
		\caption{Sub-Gradient Decent (SGD) Algorithm to Solve Problem (\ref{pro5})}
		\begin{algorithmic}\label{alg1}
			\STATE Initialize $\rho_{u,i}^{(0)}$, $\tau_{u,i}^{(0)}$, $\forall u \in \mathcal{U}$, the convergence precision $\varsigma$ and the iteration number $t=1$; 
			\REPEAT
			\STATE Calculate $\bm{\phi}^{(t)}$ according to (\ref{Phi});
			\STATE Update $\rho_{u,i}^{(t)}$ according to (\ref{rhokt});
			\UNTIL  $|\bm{\phi}^{(t)}-\bm{\phi}^{(t-1)}|\leq\varsigma$.
		\end{algorithmic}
	\end{algorithm}
	
	\begin{proposition}\label{Pro_p1}
		The SGD algorithm can find the globally optimal solution to Problem (\ref{pro5}). 
	\end{proposition}
	\textbf{Proof}: see Appendix A. 	\qed

	\subsection{Sub-band Allocation and Power Control Optimization}
	
	With the given IRS location $\bm{l}$ and the IRS coefficient $\bm{\Phi}$, the auxiliary variables, the sub-bands and the power allocation can be optimized by solving the following problem:
	\begin{subequations}\label{pro7}
		\begin{align}
		\underset{ \underset{\{t_{u,i}\}, \{p_{i}^t\},}{\{\alpha_u\}}}{\text{max}}  
		& \ R_s(t_{u,i}) =\sum_{u =1}^U\sum_{i =1}^{I} \alpha_{u,i} B_i  \log \left( 1 +  \delta_{u,i}t_{u,i}\right) 	 \label{obj7}  \\
		\text{s.t.} & \ \sum_{i =1}^I \alpha_{u,i} B_i  \log \left( 1 +  \delta_{u,i}t_{u,i}\right) \geq R_u^{th}, u \in \mathcal{U}, \label{st1}\\
		& \   p^t_i h_{u,i}^2 \geq  \alpha_{u,i} t_{u,i},\forall u \in \mathcal{U}, i \in  \mathcal{I}, \label{P7_st2}\\
		& \	(\ref{P2_st2}), C1-C2. 
		\end{align}
	\end{subequations}
	By introducing the transformation $x_{u,i} = \alpha_{u,i} t_{u,i}$, the above Problem (\ref{pro7}) can be solved by the dual-based method given in \cite{Wong.1999} and \cite{Yu.2006}.

	In summary, based on the above analysis, we propose the following BCS algorithm to solve the original problem in (\ref{pro2}), and the detailed algorithm is presented in Algorithm \ref{alg4}.
	
	\begin{algorithm}
		\caption{Block Coordinate Searching (BCS) Algorithm to Solve Problem (\ref{pro2})}
		\begin{algorithmic}[1]\label{alg4}
			\STATE Initialize searching step-sizes $\delta_x$ and $\delta_y$.
			\FOR{$X = 0, \delta_x, 2\delta_x, \cdots, \left \lfloor \frac{L}{\delta_x}\right \rfloor\delta_x$}
			\FOR{$Y = 0, \delta_y, 2\delta_y, \cdots, \left \lfloor \frac{W-(N-1)\delta}{\delta_y} \right \rfloor\delta_y$}		
			\STATE Initialize $\bm{\phi}^{(0)}$, the convergence precision $\sigma$ and the iterative number $n=0$.
			\REPEAT
			\STATE Calculate ${t_{u,i}}^{(n+1)}$, ${p^t_i}^{(n+1)}$ and ${\alpha_{u,i}}^{(n+1)}$ by solving Problem (\ref{pro7});
			\STATE  Initialize $\bm{\hat{\phi}}^{(0)} = \bm{\phi}^{(n)}$, convergence precision $\varsigma$ and iterative number $s=1$;
			\REPEAT
			\STATE Calculate $\bm{\phi}^{(s)}$ by solving Problem (\ref{pro5}) using SGD Algorithm;
			\STATE Formulate Problem (\ref{pro5}) with $\bm{\hat{\phi}}^{(s)} = \bm{\phi}^{(s)}$;
			\UNTIL  $|\bm{\phi}^{(s)} -\bm{\phi}^{(s-1)}|\leq \varsigma$;
			\STATE Set $\bm{\phi}^{(n+1)} = \bm{\phi}^{(s)}$, calculate $R^{(n+1)}$ according to (\ref{obj1});
			\UNTIL $|R^{(n+1)}-R^{(n)}|\leq \sigma$;
			\STATE Set $R(X,Y)= R^{(n+1)}$ ;
			\ENDFOR
			\ENDFOR
			\STATE  Find the optimal $(X^*, Y^*) = \arg \max R(X,Y)$;
		\end{algorithmic}
	\end{algorithm}

	In the proposed algorithm, according to \cite{Wong.1999} and \cite{Yu.2006}, the obtained solution ${p^t_i}$ and ${\alpha_{u,i}}$ satisfies the Karush–Kuhn–Tucker (KKT) conditions of Problem (\ref{pro7}). 
	Furthermore, according to \cite{Marks.1978}, by successively constructing the approximated Problem (\ref{pro5}), the converged result obtained in step 8 - step 11 is a local minimum to Problem (\ref{pro4}).
	Then, it can be verified that the objective $R^{(n)}$ is non-decreasing over each iteration, and the convergence of the procedure from  step 4 - step 14 is guaranteed.
	
	The complexity of the proposed algorithms consists of three parts: 1) The first part is the dual algorithm to find $\alpha_{u,i}$ and $p^t_i$ in step 6, of which the complexity can be concluded as $\mathcal{O}(U^4 + U^3I)$\cite{Yang.2019}; 2) The second part is the successively approximation to find $\bm \phi$. Let $S$ denote the iteration number in Step 8, and the number of iterations in SGD algorithm is $T$. Then the complexity to find $\bm \phi$ is $\mathcal{O}(ST)$; 3) The last part is the two-dimensional search to find IRS location $\bm{l}$. Let $W$ denote the iteration number in Step 5. Then, the total complexity can be concluded as $\mathcal{O}(\left \lfloor \frac{L}{\delta_x}\right \rfloor\left \lfloor \frac{W-(N-1)\delta}{\delta_y} \right \rfloor W(ST +U^4 + U^3))$.

	\section{Simulation Results}
	
	In this section, simulation results are presented to show the performance of the proposed scheme.
	UEs are uniformly distributed in a $5$m $\times$ $8$m rectangular area, the height of ceiling is $3$m, and the AP is located at $(0,0,2)$.
	The UE's rate requirement is $1$Gbps, the bandwidth of the sub-band is $50$GHz, the transmission frequency is $200$-$400$GHz, and $p_{max}$ is $1$W.
	The number of IRS reflecting elements is $20$, and the distance between each element is $5$mm.
	All the results are obtained by averaging over $100$ random realizations of UE locations.
	For comparison, we consider three different algorithms: {1) The location of the IRS is randomly chosen; 2) The phase shift of IRS is randomly generated; 3) The location of the IRS is selected to minimize the sum of  transmission distances of all UEs, i.e., $(X,Y) = \arg \underset{0 < X < W, 0 < Y < L }{\text{min}}\sum_{u  \in \mathcal{U}} d_u$.}
	The above three algorithms are labelled as ``RanLoc'',  ``RanPhi'' and ``MiniDis'', respectively.
	
	\begin{figure}
		\vspace{-1em}
			\centering
			\includegraphics[width=0.3\textwidth]{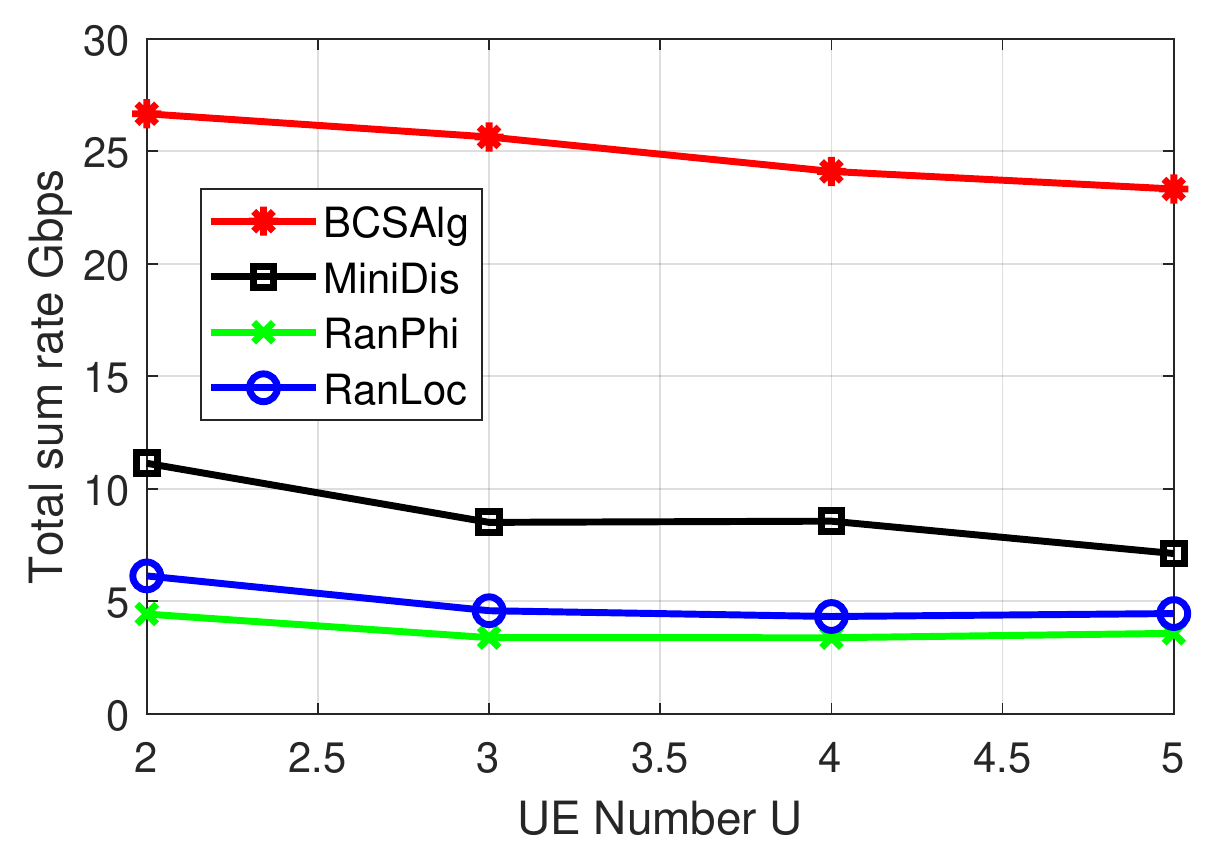}
			\vspace{-1em}
			\caption{Sum rate performance obtained by different algorithms.}
			\vspace{-2em}
			\label{fig4}
		\end{figure}
	
	Fig. \ref{fig4} illustrates the sum rate obtained by different algorithms.
	It is observed that the proposed ``BCSAlg" algorithm always achieves the best performance.
	The sum rates decrease with the number of UEs for all the algorithms, which is due to the rate requirement constraint.
	The ``MiniDis'' approach is better than the ``RanLoc'' and ``RanPhi'' schemes due to the optimized IRS location and phase shifts.
	Moreover, it is observed that the achieved rate of ``RanLoc'' is slightly larger than that of ``RanPhi''.
	 This implies that IRS phase shift has a slightly more significant impact on the sum rate performance compared with the IRS location.
	
	\section{Conclusion}
	
	In this paper, the sum rate of the UEs in an IRS-assisted THz transmission system has been maximized by optimizing the IRS location, the IRS phase shift, sub-band allocation, and power control. 
	Although the formulated problem is nonconvex, the proposed algorithm has been shown to improve the sum rate performance significantly.
	Through simulations, it is observed that our proposed solution has considerable performance gain over the benchmark schemes.

	\appendices

	\section{Proof of Proposition \ref{Pro_p1} }  \label{App1}
	\setcounter{equation}{0}   
	\renewcommand{\theequation}{\thesection.\arabic{equation}}
	
	The proposition is proved by using the contradiction method.
	
	Let $\{\rho_{u,i}^*\}$ and $\bm \phi^*(\rho_{u,i}^*)$ denote the converged results obtained by SGD algorithm.
	We define function $f_{u,i}(\bm{\phi}^*) = 2 \Re\{\bm \Theta_{u,i}(\bm{\hat{\phi}})\bm{\phi}^*\}  - \Psi_{u,i}(\bm{\hat{\phi}}) - t_{u,i}$.
	
	Assume that $\bm \phi^*(\rho_{u,i}^*)$ is not the globally optimal solution to Problem (\ref{pro5}), so that the constraint (\ref{P5_st2}) cannot be satisfied for all $u \in \mathcal{U}$.
	Let $ \mathcal{U}_1$ denotes the set of UEs that satisfy the constraint (\ref{P5_st2}), and the set of the left unsatisfied UEs are denoted by $ \mathcal{U}_2$, i.e., $ \mathcal{U} =\mathcal{U}_1 + \mathcal{U}_2$.
	Then, the globally optimal solution to Problem  (\ref{pro5}) is denoted by $\tilde{\bm \phi}$, and the following inequalities hold:
	\begin{equation}\label{contr1_1}
	\sum_{u \in \mathcal{U}_1}\sum_{i  \in \mathcal{I}_u}\rho^*_{u,i}   f_{u,i}(\bm{\phi}^*) 
	< 0 < \sum_{u \in \mathcal{U}_1}\sum_{i  \in \mathcal{I}_u}\rho^*_{u,i}  f_{u,i}(\tilde{\bm \phi}) .
	\end{equation}
	Adding the same term to both sides of  (\ref{contr1_1}), we have
	\begin{multline}\label{contr2}
	\sum_{u \in \mathcal{U}}\sum_{i  \in \mathcal{I}_u}\rho^*_{u,i}f_{u,i}(\bm{\phi}^*)
	< \sum_{u \in \mathcal{U}_1}\sum_{i  \in \mathcal{I}_u} \rho_{u,i}^*f_{u,i}(\tilde{\bm \phi})\\ + \sum_{u \in \mathcal{U}_2}\sum_{i  \in \mathcal{I}_u}\rho^*_{u,i}f_{u,i}(\bm{\phi}^*).
	\end{multline}
	Meanwhile, as the phase vector $\bm{\phi}^*$ obtained in Algorithm \ref{alg1} achieves the globally optimal solution to Problem (\ref{pro6}), we have
	\begin{equation}\label{contr3}
	\sum_{u \in \mathcal{U}}\sum_{i  \in \mathcal{I}_u}\rho^*_{u,i}f_{u,i}(\bm{\phi}^*) 
	>\sum_{u \in \mathcal{U}}\sum_{i  \in \mathcal{I}_u}\rho^*_{u,i}f_{u,i}(\tilde{\bm \phi}) .
	\end{equation}

	Then, combining the left hand side of (\ref{contr2}) and the left hand side of (\ref{contr3}),  as well as removing the common terms in $\mathcal{U}_1$, we have
	\begin{equation}\label{eqcontr}
	\sum_{u \in \mathcal{U}_2}\sum_{i  \in \mathcal{I}_u}\rho^*_{u,i}f_{u,i}(\bm{\phi}^*)> \sum_{u \in \mathcal{U}_2}\sum_{i  \in \mathcal{I}_u}\rho^*_{u,i}f_{u,i}(\tilde{\bm \phi}) .
	\end{equation}
	
	Then, we consider two cases for $\rho^*_{u,i}$: 1) $\rho^*_{u,i}=0$, $\forall u \in \mathcal{U}_2$; 2) $\rho^*_{u',i} > 0, u' \in \mathcal{U}'$, $\mathcal{U}' \subseteq \mathcal{U}_2$ and  $i \in \mathcal{I}_{u'}$.    
	
	In the first case, the left hand side and the right hand side of (\ref{eqcontr}) both equal zero, which contradicts the assumption.
	
	In the second case, as $\rho^*_{u',i} > 0$ and SGD algorithm is based on the sub-gradient method, then with a sufficient small step size, the converged result $\bm{\phi}^*$ obtained by SGD algorithm satisfies the condition of $f_{u,i}(\bm{\phi}^*)=0, \forall u' \in \mathcal{U}', i \in 
	\mathcal{I}_{u'}$.
	Then, combining with the left hand side of (\ref{eqcontr}), we have
	\begin{equation}
	0 > \sum_{u \in \mathcal{U}_2}\sum_{i  \in \mathcal{I}_u}\rho^*_{u,i}f_{u,i}(\tilde{\bm \phi}) .
	\end{equation}
	
	As $\rho_{u',i} ^*>0$, it is inferred that $f_{u,i}(\tilde{\bm \phi})  <0, \forall u' \in \mathcal{U}', i \in \mathcal{I}_{u'}$.
	This contradicts the constraints in (\ref{P5_st2}).
	However, as the $\tilde{\bm\phi}$ is the globally optimal solution to Problem  (\ref{pro5}), so that $\tilde{\bm\phi}$ should satisfy all the constraints of Problem  (\ref{pro5}).    
	As a result, the assumption does not hold, and $\bm \phi^*(\rho_{u,i}^*)$ is the globally optimal solution to Problem  (\ref{pro5}).
	Hence, the proof is completed.

	\qed

	\bibliographystyle{ieeetran}
	\bibliography{Reference}
	
\end{document}